\documentclass[a4paper,11pt]{article}

\usepackage{jheppub} 

\usepackage[T1]{fontenc} 

\usepackage{amsmath}
\usepackage{amssymb}
\usepackage{amsfonts}
\usepackage{slashed}
\usepackage{physics}
\usepackage{xcolor}
\usepackage{graphicx}
\usepackage{dcolumn}
\usepackage{bm}

\title{\boldmath Bosonization, Duality, and the $C$-Theorem in the Non-Abelian Thirring Model}

\author[a]{Rodrigo Corso B. Santos,}
\author[b]{Carlos A. Hernaski,}
\author[a]{Pedro R. S. Gomes.}

\affiliation[a]{Departamento de Física, Universidade Estadual de Londrina\\Caixa Postal 10011, 86057-970, Londrina, PR, Brasil.}
\affiliation[b]{Departamento de Física, Universidade Tecnológica Federal do Paraná, 85503-390, Pato Branco, PR, Brasil}

\emailAdd{rodrigocorso@uel.br}
\emailAdd{carloshernaski@utfpr.edu.br}
\emailAdd{pedrogomes@uel.br}

\abstract{	
	We revisit the two dimensional non-Abelian Thirring model in order to investigate its fixed point structure and the corresponding renormalization group (RG) flow. For this purpose we discuss the bosonization of the model, and we present different, but of course equivalent, bosonic versions of the theory. The bosonic theories are illuminating in that they exhibit the fixed points in a manifest way, and also possess a remarkable strong/weak duality that sheds light on the fixed point structure of the theory. We study the RG flow through the computation of the Zamolodchikov $C$-function and of the $\beta$-function in the large-level limit. Within this framework, we discuss how close to the infrared fixed point the RG flow can reach, since this point is strictly unachievable due to an emergent gauge invariance. 
}

\begin{document} 
\maketitle
\flushbottom

\section{Introduction} \label{Introduction}
The study of interacting fermions in two spacetime dimensions is absolutely important for several reasons. In  the context of high-energy physics, such systems have been vastly used as a theoretical laboratory to test a number of non-perturbative aspects, shedding light in the depths of quantum field theories \cite{Frishman:2010zz}. In the realm of condensed matter, the study of interacting fermions in one spatial dimension has led to the central concept of a Luttinger liquid, and also has been used in the description of many experimental quasi-one dimensional systems \cite{giamarchi2003quantum,gogolin2004bosonization}. 

In recent applications, fermionic one-dimensional systems have been used as the building-blocks for the construction of higher-dimensional topological phases in the framework of coupled quantum wires \cite{Kane2002,Teo2014,Iadecola_2016} (see also the review \cite{Meng:2019ket} and the references therein). The basic idea is that a higher-dimensional topological phase, say a two-dimensional one, can be seen as a set of one-dimensional quantum wires arranged in parallel. Interactions between neighboring wires allow tunneling of electrons from one wire to another, effectively realizing a two-dimensional phase. In addition to the technical advantage in treating the systems in this way, this type of approach is illuminating in that it provides a bridge between the microscopic constituents and the remarkable emergent macroscopic physics like the fractionalization of quantum numbers \cite{Fuji:2018fuj,Fontana2019,Imamura:2019uia,Toledo:2022btx}. 

With this type of application in mind, we revisit the well-studied non-Abelian Thirring model \cite{Dashen:1973nhu}, which consists of fermions belonging to the fundamental representation of a group $G$ with a current-current interaction, where we choose the currents to belong to a subgroup $H$ of $G$.  This model is used as an important ingredient (in the quantum wires sense) for the construction of a class of non-Abelian spin liquids, whose edge states support the minimal and the superconformal minimal models \cite{Huang2016,Hernaski2017}.  We would like to emphasize that, despite the fact that the non-Abelian Thirring model has been the subject of several investigations over the years, we perform an analysis that connects many distinct results within a single coherent framework, in addition to filling several gaps described below. 

In the present work we pay special attention to the bosonization of the non-Abelian Thirring model. We derive different but of course equivalent bosonized theories, which are illuminating in several respects. First, the bosonic theories enjoy a remarkable strong/weak duality \cite{Kutasov1989}. Using one of the bosonic theories we derived, we establish such duality in a  regularization-independent way, in contrast with the original derivation of \cite{Kutasov1989}.  The other form of the bosonized theory, in its turn, permits us to understand the approximate version of the duality appearing in the large-level limit, discussed in a series of works \cite{Georgiou2015,georgiou2017integrable,Georgiou2017,Georgiou2017a,Georgiou:2018vbb,hoare2019integrable,Georgiou2020}. Furthermore, upon bosonization we are able to uncover exactly the fixed points of the theory, which we connect directly with the respective values of the fermionic coupling constant.  The bosonic theory has three fixed points, but only two of them are identified as the UV and IR fixed points of the fermionic counterpart. The remaining one is not a feature of the fermionic theory, since the corresponding value of the fermionic coupling constant lies outside its domain. Furthermore, using the duality we can see that even in the bosonic theory this point seems to be problematic due to unitarity issues.

Proceeding with the bosonized theory, we study the renormalization group (RG) flow between the UV and IR fixed points through the computation of the Zamolodchikov $C$-function and of the $\beta$-function. The IR fixed point possesses a quite peculiar feature. It is strictly unachievable via renormalization group flow due to an emergent gauge invariance existing only at this point. This, in turn, implies an abrupt decoupling of degrees of freedom, and consequently a discontinuity in the $C$-function. As a consequence, the IR fixed point is infinitely far apart from the rest of the points of the parameter space \cite{Kutasov1989}.  Nevertheless, within the large-level expansion we discuss that the RG flow can reach close enough to this point by means of the zoom-in limit \cite{Sfetsos:2013wia,hoare2019integrable,Georgiou2020}. All the aspects we consider in this work are directly relevant to investigate the stability of the non-Abelian spin liquids we mentioned previously, which will be addressed in a separate publication \cite{Corso:2023}. 

This work is organized as follows: we start in Section \ref{Section2} where we discuss different ways to perform the bosonization of the non-Abelian Thirring model and how the duality transformation acts on these partition functions. Then, we proceed to Section \ref{Section3} where we use the bosonic partition functions to examine the fixed point structure of our theory. We reproduce the free fermion fixed point, and find two new ones. We follow to Section \ref{Section4}, where we discuss the relation between the fixed points in the bosonic and fermionic descriptions of the Thirring model. We find that one of the fixed points is unphysical, so that we are left only with the free fermion and an IR fixed point. In Section \ref{Section5} we examine the RG of the theory through an analysis of the Zamolodchikov metric, $ \beta$-function and $ C$-function. We find that the IR fixed point is inaccessible from the remaining RG flow. Nonetheless, we are able to probe the IR limit by taking the IR and large-$ k $ limits together. At last, in Section \ref{Section6} we present our final remarks.

\nopagebreak

\section{Non-Abelian Thirring Model} \label{Section2}

We consider the task of discussing the generating functional of current-current correlation functions of the following non-Abelian Thirring model in 2D Euclidean spacetime:
\begin{equation}
	Z[B]=\int D\psi_LD\psi_R \exp-\int \dd[2]{z} \; \left[\psi^{\dagger}_{L}\left(\partial_z-iB^a_zt^a\right)\psi_{L}+\psi^{\dagger}_{R}\left(\partial_{\bar{z}}-iB^a_{\bar{z}}t^a\right)\psi_{R} +\frac{\lambda}{k} J_{R}^{a}J_{L}^{a}\right],
	\label{5.1}
\end{equation}
with $\partial_{z}\equiv\partial_{\tau}-i \partial_{1} $ and $\partial_{\bar z}\equiv\partial_{\tau}+i\partial_{1}$, and the spinors
belong to the fundamental representation of a group $G$. The objects $J_{R/L}$ are the fermionic currents of a subgroup $H$ of $G$:
\begin{eqnarray}
	J^{a}_{R/L}&=& \psi_{R/L}^{\dagger}t^{a}\psi_{R/L},\label{5.2}
\end{eqnarray}
where the $t^a$ are the generators of $H$ in the fundamental representation of $G$. Let $t^a_F$ be generators of the subgroup $H$ in its own fundamental representation. We adopt the normalization of these generators to be
\begin{equation}
	\tr\left(t^{a}_F t^{b}_F\right)=\frac{1}{2}\delta^{a b}.\label{2.3}
\end{equation}
The parameter $k$ is the index of the fermion representation: a positive integer defined by $\tr(t^at^b)=k\tr(t^a_{F}t^b_{F})$.

One of our main goals is to investigate the existence of a non-trivial fixed point in the IR limit, with central charge corresponding to the coset structure $G/H$. For $\lambda=0$, the theory (\ref{5.1}) consists of a conformal model of a set of free fermions. According to the Sugawara construction, the corresponding energy-momentum tensor can be decomposed as
\begin{align}
	T_G=T_{G/H}+T_H,\label{5.4a}
\end{align}
which implies the splitting of the central charge 
\begin{align}
	c_G=c_{G/H}+c_{H}. \label{full_central_charge}
\end{align}
If we choose the sign of $\lambda$ to be positive, the four fermion current-current interaction turns out to be relevant \cite{tsvelik2015lectures}. Therefore, turning on this relevant operator, the degrees of freedom partaking in the interaction open a mass gap and we end up with the conformal model with a coset $G/H$ structure in the IR regime.


In the following sections, we investigate this scenario by calculating the $\beta$-function of $\lambda$ up to the order of $1/k$ in a large-$ k $ expansion and use the $C$-theorem to explicitly construct the $C$-function that interpolates between the UV and IR fixed points.


\subsection{Bosonization}

In this section we discuss how to get bosonized versions of (\ref{5.1}). Besides the fixed point structure be significantly more transparent in terms of bosonic variables, we will be able to show that there is an interesting duality in the Thirring model, which was already investigated in \cite{Kutasov1989}, and will be useful in the investigation of $\beta$- and $C$-functions.

In order to bosonize the model (\ref{5.1}), we first rewrite the generating functional by using auxiliary vector fields valued in the algebra of the $H$-subgroup. Then, we get
\begin{align}
	Z[B]=\int DAD\psi\exp-\int \dd[2]{z} \; & \Big[\psi^{\ast}_{L}\left(\partial_{z}-i\left(A^a_z+B^a_z\right)t^a_R\right)\psi_{L} +\psi^{\ast}_{R}\left(\partial_{\bar{z}}-i\left(A^a_{\bar{z}}+B^a_{\bar{z}}\right)t^a_R\right)\psi_{R} \nonumber \\
	&\left.+\frac{2k}{\lambda}A^a_{z}A^a_{\bar{z}}\right ],
	\label{5.3}
\end{align}
so that we recover (\ref{5.1}) upon the integration of the auxiliary fields. We notice that, since $k/\lambda$ is positive, the gaussian integration actually converges. Furthermore, for the limit $\lambda\rightarrow 0$ to be well-defined and commute with the integration of the auxiliary fields, we need to impose that $A_{z}=A_{\bar{z}}=0$ in this limit. 

Redefining $A\rightarrow A-B$ and integrating the fermion fields we obtain   
\begin{align}
	Z[B]=\int DA\;Z[A]\exp\int \dd[2]{z}\frac{2k}{\lambda}\text{Tr}\left(A_{z}-B_z\right)\left(A_{\bar{z}}-B_{\bar{z}}\right), \label{5.4} 
\end{align}
where $A\equiv -iA^at^a$, $B\equiv -iB^at^a$, and
\begin{eqnarray}
	Z[A]=\det\left(D_z\right)\det\left(D_{\bar{z}}\right). \label{5.5}
\end{eqnarray}
The determinants of the covariant derivatives can be explicitly calculated \cite{Polyakov:1983tt}:
\begin{eqnarray}
	Z[A]=Z_F\exp k\left[W[M^\dagger]+W[M]+\frac{b}{\pi}\int \dd[2]{z} \Tr \left(A_{\bar{z}}A_z\right)\right]\label{5.6}
\end{eqnarray}
where $W[M]$ is the Wess-Zumino-Witten (WZW) action
\begin{align}
	W[M]&=\frac{1}{2\pi}\int_{\partial \mathcal{M}} \dd[2]{z}\text{Tr}\left(\partial_{z}M\partial_{\bar{z}}M^{-1}\right)\nonumber\\
	&+\frac{i}{12\pi}\int_{\mathcal{M}} \dd[3]{z} \epsilon^{\mu\nu\sigma}\text{Tr}\left(\partial_{\mu}M M^{-1}\partial_{\nu}M M^{-1}\partial_{\sigma}MM^{-1}\right).\label{2.22}
\end{align}
In (\ref{5.6}), $Z_F=\det(\partial_z)\det(\partial_{\bar{z}})$ is the partition funtion for a set of free fermions, and we have used the parametrizations 
\begin{equation}
	A_z=\partial_z M M^{-1}\ \  \text{and}\ \  A_{\bar{z}}=-M^{\dagger -1}\partial_{\bar{z}} M^{\dagger},\label{5.7}
\end{equation}
with $M$ and $M^{\dagger}$ being complex $H$-valued matrices. The parameter $b$ incorporates the regularization ambiguities in the bosonization procedure. 

When we have fermions minimally coupled to gauge fields, it is natural to choose $b=1$, which ensures the vector gauge invariance of the fermionic model. However, in the present case the fermion determinant $Z[A]$ appears only as a piece of the complete Thirring model, which does not exhibit the mentioned gauge invariance due to the quadratic term in the gauge field appearing in (\ref{5.4}). Choosing $b=1$ in \eqref{5.6} amounts to ensuring that \eqref{5.3} has an emergent gauge invariance only for $\lambda\rightarrow\infty$. A deviation from $b=1$ can be absorbed in the $\lambda$ parameter causing a shift in the value at which the gauge invariance arises without affecting the physics of the model. In other words, the specific value of the fixed points of the theory are $b$-dependent (i.e., regularization-dependent), but their existence is  independent of the regularization choice. Nevertheless, as we shall see, to obtain the different bosonized versions of the Thirring model we need to perform Gaussian integrations whose convergences impose further restrictions on $ b $.

The generating functional (\ref{5.4}) together with the expression (\ref{5.6}) for the fermion determinants, and a further change of variables from $A_{z}$ and $A_{\bar{z}}$ to $M$ and $M^{\dagger}$, can be seen as a bosonized version of the Thirring model \eqref{5.1}. We will return to this point later when we discuss duality.

To proceed, we consider the following identities:
\begin{align}
	W[M^\dagger g M]&-W[M^\dagger]-W[M]=W[g]\nonumber \\
	&-\frac{1}{\pi}\int d^2z \Tr\left(A_{z}g^{-1}\partial_{\bar{z}}g-A_{\bar{z}} \partial_z gg^{-1}-A_{\bar{z}} g A_{z} g^{-1}\right)\label{5.8}
\end{align}
and
\begin{align}
	W[M^\dagger g]+W[g^\dagger M]&-W[M^\dagger]-W[M]=W[g]+W[g^\dagger]\nonumber\\
	&-\frac{1}{\pi}\int d^2z \tr\left(A_{z}g^{\dagger-1}\partial_{\bar{z}}g^\dagger-A_{\bar{z}} \partial_z gg^{-1}\right),\label{5.9}
\end{align}
with $A_{z}$ and $A_{\bar{z}}$ given by (\ref{5.7}).
These relations follow directly from the application of the Polyakov-Wiegmann identity
\begin{eqnarray}
	W[gh]=W[g]+W[h]-\frac{1}{\pi}\int d^2z \Tr\left(g^{-1}\partial_{\bar{z}}g\partial_{z}hh^{-1}\right).\label{5.10}
\end{eqnarray}

We can use the identities \eqref{5.8} and \eqref{5.9} to rewrite the $Z[A]$ in (\ref{5.6}) in two distinct but equivalent forms. In the first one, we multiply and divide the RHS of (\ref{5.6}) by 
\begin{align}
	Z_k=\int Dg e^{-kW[g]} \label{5.10.1}
\end{align}
to get
\begin{eqnarray}
	Z[A]=\frac{Z_F}{Z_k}\int Dg e^{-k\left(W[g]-W[M^\dagger]-W[M]-\frac{b}{\pi}\int \dd[2]{z} \Tr \left(A_{\bar{z}}A_z\right)\right)}.\label{5.11}
\end{eqnarray}
Performing the change of variable $g$ $\rightarrow$ $M^\dagger g M$, which has unit Jacobian by the invariance of the Haar measure, and using the identity (\ref{5.8}), we find
\begin{eqnarray}
	Z^{(b)}_1[A]=\frac{Z_F}{Z_k}\int Dg e^{-k\left(W[g]-\frac{1}{\pi}\int d^2z \Tr\left(A_{z}g^{-1}\partial_{\bar{z}}g-A_{\bar{z}} \partial_z gg^{-1}-A_{\bar{z}} g A_{z} g^{-1}+bA_{\bar{z}}A_z\right)\right)}.\label{5.12}
\end{eqnarray}

Alternatively, we can multiply and divide (\ref{5.6}) by $Z_k^2$ given by \eqref{5.10.1}, perform the changes $g$ $\rightarrow$ $g M$ and $g^\dagger$ $\rightarrow$ $M^\dagger g^\dagger$, and use the identity (\ref{5.9}) to obtain
\begin{align}
	Z^{(b)}_2[A]=\frac{Z_F}{Z_k^2}\int DgDg^\dagger e^{-k\left(W[g]+W[g^\dagger]-\frac{1}{\pi}\int d^2z \Tr\left(A_{z}g^{\dagger-1}\partial_{\bar{z}}g^\dagger-A_{\bar{z}} \partial_z gg^{-1}+bA_{\bar{z}}A_z\right)\right)}.\label{5.13}
\end{align}
In spite of using the labels ``1'' and ``2'' to identify these generating functionals, we emphasize that they are equivalent.

If we insert $Z^{(b)}_{1}$ in (\ref{5.4}) and integrate in the auxiliary gauge fields, we have 
\begin{align}
	Z[B]&=\frac{Z_F}{Z_k}\int Dg\frac{1}{\det k(\xi^{-1}-D^T)} \exp-k\left\{ W[g]+\int d^2z\left[\frac{\pi}{2}J^a_R\left(\xi^{-1}-D^T\right)^{-1 ab}J^b_L\right.\right.\nonumber\\
	&\left.\left.-\frac{i\left(\xi^{-1}-b\right)}{2}\left(B^a_{\bar{z}}\left(\xi^{-1}-D^T\right)^{-1 ab}J^b_R+B^a_z\left(\xi^{-1}-D^T\right)^{-1 ab}J^b_R\right)\right.\right.\nonumber\\
	&\left.\left.+\frac{\xi^{-1}-b}{2\pi}B^a_{\bar{z}}\left(\delta^{ab}-\left(\xi^{-1}-b\right)\left(\xi^{-1}-D^T\right)^{-1 ab}\right)B^b_z\right]\right\}, \label{5.14}
\end{align}
with the currents
\begin{align}
	J_L&=-\frac{1}{\pi}g^{-1}\partial_{\bar{z}}g\qquad \text{and}\qquad  J_R=\frac{1}{\pi}\partial_{z}gg^{-1},\label{5.15} 
\end{align}
and
\begin{align}
	D^{ab}&=2\Tr\left(t^agt^bg^{-1}\right), \label{5.16.0} \\
	\xi&=\frac{\lambda}{\lambda b+2\pi}.\label{5.16}
\end{align}
We notice that for the integration leading from (\ref{5.4}) to (\ref{5.14}) to make sense, we need to guarantee that the Gaussian integral converges. Since $D(g)$ is an orthogonal matrix, its eigenvalues all have unit modulus, and thus, we need that $\xi^{-1}\geq1$, or $ 0\leq \xi \leq 1$. The relation \eqref{5.16} allows to express this restriction in terms of $\lambda$, namely, 
\begin{align}
-\frac{2\pi}{b} \leq \lambda\leq \frac{2\pi}{1-b},
\label{cond}
\end{align}
which implies that $0 \leq b \leq 1$, with positive $\lambda$.

The model (\ref{5.14}) is a bosonized version of Thirring model that enables us to make contact with some works in the literature. Indeed, for small $\xi$ it is in the class of deformed sigma models, which are extensively studied in the Refs. \cite{Sfetsos:2013wia,Georgiou2015,georgiou2017integrable,Georgiou2017,Georgiou2017a,Georgiou:2018vbb,hoare2019integrable,Georgiou2020}. In those works, the authors argue that (\ref{5.14}) without the determinant in the denominator incorporates the quantum corrections of the order $1/k$ in a large-$k$ expansion of the Thirring model. We will discuss this point in the next section. 

An alternative bosonized version is obtained using $Z_2$ in (\ref{5.4}) and integrating in the gauge fields:
\begin{align}
	Z[B]&=\frac{Z_F}{Z_k^2}\int DgDg^\dagger \exp-k\left(W[g]+W[g^\dagger]+\int d^2z\left[\frac{\pi}{2}\xi J^{a}_R\left(g\right)J^{ a}_L\left(g^\dagger\right)\right.\right.\nonumber\\
	&+\frac{i\left(b\xi-1\right)}{2}\left(B^a_{\bar{z}}J^{a}_R\left(g\right)+B^a_{z}J^{ a}_L\left(g^\dagger\right)\right)+\left.\left.\frac{b\left(1-b\xi\right)}{2\pi}B^a_{\bar{z}}B^a_z\right]\right).\label{5.18}
\end{align}
In this case, the condition for the Gaussian integration convergence is $\xi\geq 0$, which in terms of $\lambda$ implies $-\frac{2\pi}{b} \leq \lambda$. In turn, this corresponds to the lower bound of the condition \eqref{cond} following from the duality \eqref{5.14}. Therefore, the partition functions \eqref{5.14} and \eqref{5.18} are valid bosonized versions of the Thirring model provided that $0 \leq b \leq 1$.

We would like to emphasize the difference between the bosonized versions of the Thirring model expressed by (\ref{5.14}) and (\ref{5.18}). In (\ref{5.14}) we have just one WZW model deformed by a highly non-linear interaction, whereas in (\ref{5.18}) there are two independent WZW models with an interaction of the same type as the original fermionic model.


\subsection{Duality}

One of the great advantages of bosonizing the Thirring model is related to the presence of an interesting duality in the theory, which in its bosonized version can be explicitly derived.

This can be attained by considering (\ref{5.4}) with the explicit form of $Z[A]$, given in (\ref{5.6}), and a further change of variables from $A_{z}$ and $A_{\bar{z}}$ to $M$ and $M^\dagger$ given by (\ref{5.7}). As is well known, the change of variables generates a non-trivial Jacobian given by a WZW term \cite{polyakov1984goldstone,Cabra1990}
\begin{eqnarray}
	DA_zDA_{\bar{z}}=\det\left(\partial_z\partial_{\bar{z}}\right)_{adj}e^{2C_H\left(W[M^\dagger]+W[M]\right)}DMDM^\dagger,\label{6.1}
\end{eqnarray} 
where $C_H$ is the quadratic Casimir invariant of the subgroup $H$ in the adjoint representation. Possible regularization ambiguities are incorporated in the $b$ parameter in \eqref{5.6}.  The subscript $adj$ of $\det\left(\partial_z\partial_{\bar{z}}\right)_{adj}$ stands for the adjoint representation  of the $H$-subgroup that the vector fields belong. Furthermore, in \cite{Kutasov1989} it is shown that the change of variables needs a further renormalization factor $f(k)$ for the current coupling to $A_{z}$ and $A_{\bar{z}}$. This factor is crucial for the presence of the duality. Therefore, we obtain 
\begin{align}
	Z[B]&=Z_FZ_g\int DMDM^\dagger e^{-\left(k+2C_H\right)\left(-W[M^\dagger]-W[M]\right)-\frac{k\xi^{-1}}{\pi}\left(f(k)\right)^2\int \dd[2]{z} \Tr\left(M^{\dagger -1}\partial_{\bar{z}}M^\dagger \partial_zMM^{-1}\right)}\nonumber\\
	&\times e^{\int \dd[2]{z}\frac{kf(k)\left(\xi^{-1}-b\right)}{\pi}\Tr\left(B_zM^{\dagger-1}\partial_{\bar{z}}M^\dagger-B_{\bar{z}}\partial_zMM^{-1}\right)-\frac{k\left(\xi^{-1}-b\right)}{2\pi}B^a_{z}B^a_{\bar{z}}},\label{6.2}
\end{align}
where $Z_g=\det\left(\partial_z\partial_{\bar{z}}\right)_{adj}$ is a partition function that can be calculated in terms of an action for ghost fields, given by
\begin{equation}
	S_{ghost}=\int \dd[2]{z}\sum\limits_{i=1}^{d_H} \left(b^{i}_{z}\partial_{\bar{z}}\bar{c}^{i}+b^{i}_{\bar{z}}\partial_{z}c^{i}_{z}\right),
	\label{2.27}
\end{equation}
where $d_H$ is the dimension of the subgroup $H$. The ghost fields $b^i_{z}$ and $b^i_{\bar{z}}$ have conformal weight one, while the fields $c^i$ and $\bar{c}^i$ have conformal weight zero.

In \cite{Kutasov1989}, the explicit expression $f(k)=\sqrt{\frac{k+2C_H}{k}}$ is obtained using a left-right symmetric regularization in the functional integral manipulations. Here, we show that the same result holds in any regularization. In fact, using $f(k)=\sqrt{\frac{k+2C_H}{k}}$ and identifying $M$ and $M^\dagger$ with $g$ and $g^\dagger$, respectively, we get 
\begin{align}
	Z[B]&=Z_FZ_g\int DgDg^\dagger \exp \left(k+2C_H\right)\left(W[g]+W[g^\dagger]+\frac{\pi}{2}\xi^{-1}\int d^2zJ^{a}_R\left(g\right)J^{ a}_L\left(g^\dagger\right)\right)\times\nonumber\\
	&\times\exp \left(k+2C_H\right)\left[i\frac{\left(\xi^{-1}-b\right)}{2}\sqrt{\frac{k}{k+2C_H}}\left(B^a_{\bar{z}}J^{a}_R\left(g\right)+B^a_zJ^{ a}_L\left(g^\dagger\right)\right)\right. \nonumber\\
	&\hspace{5.9cm}-\left.\frac{\left(\xi^{-1}-b\right)}{2\pi}\sqrt{\frac{k}{k+2C_H}}B^a_{z}B^a_{\bar{z}}\right].\label{6.3}
\end{align}
Comparing (\ref{5.18}) with (\ref{6.3}), we obtain the following duality relation between the two models: 
\begin{eqnarray}
	k &\rightarrow& -\tilde{k}=-k-2C_H,\label{6.4}\\
	\xi&\rightarrow& \tilde{\xi}=\xi^{-1},\label{6.5}\\
	B^a&\rightarrow& \tilde{B}^a=-\frac{1}{\xi}\sqrt{\frac{k}{k+2C_H}}B^a. 
	\label{6.6}
\end{eqnarray}

It is illuminating to compare our results with other similar discussions in the literature \cite{Georgiou2017,georgiou2017integrable,Georgiou2017a,Georgiou2020,hoare2019integrable,Georgiou2015}. In particular, in \cite{georgiou2017integrable}, the authors consider the following model
\begin{eqnarray}
	S^{\left(k,\xi\right)}&=&k\left(W[g]+\frac{\pi\xi}{2}\int d^2zJ^a_RJ^b_L\right)\label{6.61}
\end{eqnarray}
as the bosonized version of the Thirring model, and conjecture that 
\begin{eqnarray}
	S^{\left(k,\xi\right)}_{eff}=k\left(W[g]+\frac{\pi}{2}\int d^2zJ^a_R\left(\xi^{-1}-D^T\right)^{-1 ab}J^b_L\right)\label{6.7}
\end{eqnarray}
is an effective action of the model \eqref{6.61} incorporating the contributions to all orders in the interaction parameter but to leading order in $1/k$ expansion. In fact, one can show the identity
\begin{eqnarray}
	S^{\left(-k,\xi^{-1}\right)}_{eff}[g^{-1}]=S_{eff}^{\left(k,\xi\right)}[g].\label{6.9}
\end{eqnarray}
This gives rise to the duality: $k\rightarrow -k$ and $\xi\rightarrow \xi^{-1}$, which is the large-$k$ limit of the duality relations (\ref{6.4}) and (\ref{6.5}).  Taking $B=0$ in \eqref{5.14} and \eqref{5.18}, we notice that the model \eqref{6.61} differs from both of our bosonized actions. In fact, we could obtain \eqref{6.61} directly from the fermionic Thirring model by using the non-Abelian bosonization rules for {\it free} fermionic currents \cite{Witten1984}. In this sense, \eqref{6.61} is justified perturbatively in $\xi$, which can also be obtained by expanding \eqref{6.7} to first order in $\xi$. This is also consistent with our results, since in the large-$k$ limit we can discard the determinant factor in the denominator of (\ref{5.14}), resulting in the action (\ref{6.7}) and showing that this is a large-$k$ bosonic effective action for the Thirring model.

In the following sections, the duality relations will be helpful to understand the fixed point structure of the model as well as the analysis of the $\beta$- and $C$-functions.


\section{Fixed Points}\label{Section3}

By making $B=0$ in (\ref{5.18}) and (\ref{6.3}) it is straightforward to verify which values of $\xi$  lead to a conformal field theory. These points are given by
\begin{align}
	\xi=0 \ \ \ \ &\longleftrightarrow\ \ \ \ \ \tilde{\xi}\rightarrow \infty;\label{7.1} \\
	\xi=1\ \ \ \ &\longleftrightarrow\ \ \ \ \ \tilde{\xi}=1;\label{7.2} \\
	\xi\rightarrow\infty\ \ \ \ &\longleftrightarrow\ \ \ \ \ \tilde{\xi}=0.\label{7.3}
\end{align}
One important check of the duality relations is the comparison between the central charges of the dual models. We use the notation $c_{\xi}$ and $\tilde{c}_{\tilde{\xi}}$ for the central charges of the original and dual model, respectively. The partition functions, whose central charges we will calculate, will be formed from: free fermions, $Z_F$; ghosts, $Z_g$; and level-$k$ WZW partition functions, $Z_k$; with corresponding central charges
\begin{align}
	c_F &= N;\label{7.31}\\
	c_g&=-2d_H;\label{7.32}\\
	c_k&= \frac{kd_H}{k+C_H}.	\label{7.33}
\end{align}
Let us analyse each one of the fixed points (\ref{7.1})-(\ref{7.3}) below. 

\subsubsection{$\xi=0 \ \ \ \ \longleftrightarrow\ \ \ \ \ \tilde{\xi}\rightarrow \infty$}

For $\xi=0$, the two WZW actions $W[g]$ and $W[g^\dagger]$ decouple and then the model is conformal. The central charge can be calculated from the sum of the central charges of all decoupled terms. Therefore, the total partition function is given by $Z=\frac{Z_FZ^2_k}{Z^2_k}=Z_F$, with total central charge
\begin{align}
	c_0&=N-2\frac{kd_H}{k+C_H}+2\frac{kd_H}{k+C_H}\nonumber\\
	&=N.\label{7.4}
\end{align}
In the first line we have explicited all the central charges for the separated pieces using the expressions \eqref{7.31} and \eqref{7.33}. The negative central charge corresponds to the two WZW models in the denominator of (\ref{5.18}), whereas the last term corresponds to the two decoupled WZW models in the numerator. 

In the dual model, this point corresponds to $\tilde{\xi}=\infty$ in (\ref{6.3}). Following the discussions below equation \eqref{5.3}, this limit produces the delta functions
\begin{align}
	\lim \limits_{\tilde \xi\rightarrow \infty}  e^{-\frac{\tilde \xi}{2\pi} \int \dd[2]{z} \Tr g^{\dagger -1}\partial_{\bar z}g^{\dagger} \partial_{z}g g^{-1} }\sim	\delta\left(\partial_zgg^{-1}\right)\delta\left(g^{\dagger-1}\partial_{\bar{z}}g^\dagger\right)&=\frac{\delta\left(g-1\right)\delta\left(g^\dagger-1\right)}{\det\left(\partial_z\partial_{\bar{z}}\right)_{adj}}.\label{7.5}
\end{align}
This identity in (\ref{6.3}) cancels the WZW terms and the ghost partition function $Z_g$, which is the determinant of the partial derivatives in (\ref{6.1}). Therefore, for the dual model we only have the contribution of the free fermion $Z_F$:
\begin{align}
	\tilde{c}_{\infty}&=N\label{7.6},
\end{align}
agreeing with (\ref{7.4}).

\subsubsection{$\xi=1 \ \ \ \ \longleftrightarrow\ \ \ \ \ \tilde{\xi}=1$}

For $\xi=1$, we can use the Polyakov-Wiegmann identity (\ref{5.10}) to rewrite the two WZW models as $W[g^\dagger g]$. Making the change of variables $g^\dagger g\rightarrow g$ implies $\int DgDg^\dagger=\int D\left(g^\dagger g\right)Dg^\dagger$. The decoupled $g^\dagger$ integration gives rise to an infinite gauge volume, which is due to an emergent gauge invariance at this point:
\begin{eqnarray}
	g^\dagger\rightarrow g^{\dagger\prime}=g^\dagger\Lambda^{-1}\left(z,\bar{z}\right)\ \ \ \text{and}\ \ \ \ g\rightarrow g'=\Lambda\left(z,\bar{z}\right)g,\label{7.71}
\end{eqnarray}
with $\Lambda\in H$. Note that this is a peculiar feature of the model. This emergent gauge invariance makes some degrees of freedom suddenly decouple from the spectrum. In this way, the Zamolodchikov $ C $-function that we shall compute shortly, and which quantifies the number of degrees of freedom, is expected to be discontinuous. Fixing the gauge $g^\dagger=1$, one finally obtains the factorized partition function $Z=\frac{Z_F}{Z^2_k}Z_k=\frac{Z_F}{Z_k}$, with the central charge
\begin{eqnarray}
	c_1=N-\frac{kd_H}{k+C_H},\label{7.8}
\end{eqnarray}
which corresponds to the central charge for the coset structure $G/H$.
For the dual model, we also have $\tilde{\xi}=1$ in (\ref{6.3}). Following a similar sequence of steps we did above, we get $Z=Z_FZ_gZ_{-k-2c}$, with central charge 
\begin{align}
	\tilde{c}_1&=N-2d_H+\frac{\left(-k-2C_H\right)d_H}{(-k-2C_H)+C_H},\nonumber\\
	&=N-\frac{kd_H}{k+C_H},\label{7.9}
\end{align}
which is the same as (\ref{7.8}).

\subsubsection{$\xi\rightarrow \infty \ \ \ \ \longleftrightarrow\ \ \ \ \ \tilde{\xi}=0$}

We may wonder if $ \xi\rightarrow \infty $ corresponds to a fixed point. It is difficult to analyze this limit directly in \eqref{5.18}, as it seems to introduce divergences in the partition function. Nonetheless, we are able to consider this limit by writing the partition function \eqref{5.18} before the integration over the fields $ A_{z}$ and $ A_{\bar z}$:
\begin{align}
	\lim \limits_{\xi \rightarrow \infty}&\frac{Z_F}{Z_k^2}\int DgDg^\dagger DA e^{-k\left[W[g]+W[g^\dagger]-\frac{1}{\pi}\int d^2z \Tr\left(A_{z}g^{\dagger-1}\partial_{\bar{z}}g^\dagger-A_{\bar{z}} \partial_z gg^{-1}+\xi^{-1}A_{\bar{z}}A_z\right)\right]}\nonumber \\
	=&\frac{Z_F}{Z_k^2}\int DgDg^\dagger e^{-k\left(W[g]+W[g^\dagger]\right)} \delta(\partial_{ z}gg^{-1}) \delta(g^{\dagger -1}\partial_{\bar z}g^\dagger).
\end{align}
In this case, the analysis is similar to what we did for the $\tilde{\xi}\rightarrow \infty$ in the dual model. Again we get constraints making the currents to vanish, and using (\ref{7.5}) in the above expression we get $Z=\frac{Z_F}{Z^2_kZ_g}$, which yields
\begin{align}
	c_\infty&=N-2\frac{kd_H}{k+C_H}+2d_H\nonumber\\
	&=N+\frac{2C_Hd_H}{k+C_H}.\label{7.10}
\end{align}
In the dual model we take $\tilde{\xi}=0$ to get $Z=Z_FZ_gZ^2_{-k-2C_H}$ with the same central charge
\begin{align}
	\tilde{c}_0&=N+2\frac{\left(k+2C_H\right)d_H}{k+C_H}-2d_H\nonumber\\
	&=N+\frac{2C_Hd_H}{k+C_H}.\label{7.10}
\end{align}

We would like to emphasize that, in spite of the fact that a conformal field theory seems to emerge at $  \xi\rightarrow \infty$, the theory at this point (actually in the entire strong coupling region) has the potential to exhibit unphysical properties since it is dual to a theory with negative levels at weak coupling. Furthermore, as we shall see in the next section, this fixed point corresponds to a negative value of the fermionic coupling constant $ \lambda $, which lies outside the allowed values. We will also see that the problematic strong coupling region $ \xi>1 $ is safely inaccessible from $ 0\leq \xi <1 $, which corresponds to the range of interest of the original fermionic model. Therefore, our analysis in the remaining of this work will be concerned with the region $ 0\leq \xi <1 $.


\section{Fixed Points Structure and Duality in the Fermionic Theory} \label{Section4}

It is interesting to discuss the relation of these fixed points of the bosonic theories with the corresponding fixed points in the fermionic Thirring model. As we already pointed out, the comparison of the fermionic and bosonic models should be made with care due to regularization ambiguities and convergence issues of functional integrations. To this end, let us express the duality in terms of the fermionic parameters. First, we define $\tilde{\lambda}$ for the dual model of (\ref{5.1}) through the relation analogue to (\ref{5.16}):
\begin{eqnarray}
	\tilde{\xi}&=&\frac{\tilde{\lambda}}{\tilde{\lambda} b+2\pi}.\label{7.11}
\end{eqnarray}
From (\ref{6.5}) we can express the relation between the original parameter $\lambda$ of the current-current interaction in the Thirring model (\ref{5.1}) and $\tilde{\lambda}$:
\begin{equation}
	\tilde{\lambda}=\frac{2\pi\left(2\pi+b\lambda\right)}{\lambda\left(1-b^2\right)-2\pi b}.\label{7.12}
\end{equation}
According to the discussion below \eqref{5.7}, the original Thirring model does not possess any gauge invariance for generic values of $\lambda$, and one can argue that there is no preferred choice for the regularization parameter in the calculation of the fermion determinant. However, in the light of our discussion, it is possible to explore the constraints on the parameters of the theory due to the consistency of the duality web relating the fermionic and bosonic  models. 

We will focus on the bosonic version (\ref{5.18}) and its dual model (\ref{6.3}). We already obtained the constraint  $0 \leq b \leq 1 $ as a consistency condition for these bosonic models to describe the fermionic counterpart. Furthermore, the model possesses an emergent gauge invariance as $\lambda\rightarrow\infty$ if we choose $b=1$ in \eqref{5.6}. A different choice for $b$ only hides this fact by shifting the specific value of $\lambda$ where this gauge redundancy emerges. Therefore, we will take $b=1$ for simplicity. With this, all the range of $0\leq\lambda<\infty$ of the fermionic model is contained within the range $0\leq\xi<1$ of its bosonized version (\ref{5.18}). In terms of dual coupling $\tilde{\lambda}$, the relation (\ref{7.12}) gives
\begin{equation}
	\tilde{\lambda}=-\left(2\pi+\lambda\right).\label{7.13}
\end{equation}
The range $0\leq\lambda<\infty$ then implies $-\infty<\tilde{\lambda}\leq -2\pi$ for the dual fermionic model. This, in turn, is contained within the range $1<\tilde{\xi}<\infty$ of the dual bosonic one (\ref{6.3}).

Choosing $b=1$ and using the relations (\ref{5.16}) and (\ref{7.11}) together with (\ref{7.1})-(\ref{7.3}) we have the relations, expressed in table (\ref{Table1}), between the parameters of the models in the fixed points analysed above.
\begin{table}[h!] \centering
	\begin{tabular}{|c|c|c|c|}
		\hline 
		$\xi$ &	$\lambda$ & $\tilde{\xi}$ & $\tilde{\lambda}$   
		\\
		\hline 
		$0^{+}$ & $0^{+}$ & $\infty$ & $-2\pi^{-}$ 
		\\ 
		\hline
		$1^{-}$  & $\infty$ & $1^{+}$ &  $-\infty$
		\\ 
		\hline 
		$\infty$  & $-2\pi$ & $0$ & $0$ 
		\\ 
		\hline 
	\end{tabular} 
	\caption{Fixed points in the non-Abelian Thirring model.}
	\label{Table1}
\end{table}

From this table, we can also discuss the results of \cite{Dashen:1973nhu}. Through an operator analysis the authors discuss the existence of possible conformal fixed points in the fermionic Thirring model. Besides the trivial fixed point corresponding to free fermions, they find another conformal point at a non-vanishing coupling constant. However, as it has been observed in \cite{Schroer:1974zk,DellAntonio1975,Gomes:1987fm}, this last fixed point is simply a different parametrization of the trivial one, which can be interpreted as a dual description of free fermions. This fact matches with the duality in the bosonic theories, which was used to find the relation \eqref{7.13}, implying a nonvanishing value of $\tilde{\lambda}$ when $\lambda=0$.

Using the conventions of \cite{Dashen:1973nhu}, the value of the coupling constant is $2\pi$, which is the same as the absolute value of the dual of the free fixed point in table \ref{Table1}. The sign difference concerns their imposition of positive energy-momentum tensor and was already noticed in \cite{Karabali:1988sz}, whose work also found this result. However, in \cite{Karabali:1988sz}, the duality relation considers $k\rightarrow -k$, since they use a bosonic model analogue to (\ref{6.7}), which is only valid for large $k$. The last fixed point in the table also deserves some comments. Notice that it implies a negative value for the fermionic coupling constant $ \lambda $, which lies outside the domain of the fermionic Thirring model. Furthermore, even from the bosonic theory this seems to be a problematic fixed point according to the discussion in the end of the previous section and will be discarded in the following. To summarize, the bosonic versions describe the physics of the fermionic Thirring model from the UV free fermion fixed point $ (\xi=0) $ to the IR one $ (\xi=1) $ with central charge corresponding to the coset structure $G/H$.


\section{Large-level Expansion and $C$-function} \label{Section5}

The Zamolodchikov $C$-theorem \cite{Zomolodchikov1986} is one of the most precious gems of two-dimensional quantum field theories. The theorem ensures the existence of a function $C$ of the coupling constants, which is decreasing along the flow of the renormalization group and, at the fixed points, its value coincides with the central charge of the system  $C(\lambda^*)=c$. It is therefore a formalization of the intuitive notion that as we go from higher to lower energies we are integrating out degrees of freedom and hence losing information.

The $C$-function is defined as
\begin{align}
	C=(2\pi)^{2}\left[2 x^{4} \expval{T(x)T(0)}-x^{3 }\bar{x}\expval{T(x)\Theta(0)}-\frac{3}{8}x^{2}\bar{x}^{2}\expval{\Theta(x)\Theta(0)}\right], 
	\label{3.0}
\end{align}
where $T$ and $\Theta$ are combinations of the components of the energy-momentum tensor, namely, $ T(x)\equiv T_{zz}(x)$ and $ \Theta(x)\equiv \eta^{\mu \nu}T_{\mu \nu}(x)$.

In this section we assume $G=U(N)$ and $H$ is a subgroup $SU(N_c)$, with $N_c$ being the number of colors of each of the $N_f=N/N_c$ fermions. We will compute the $C$-function for the Thirring model using its bosonic version (\ref{6.3}) considering the RG flow between the fixed points $\tilde{\xi}=1$ and $\tilde{\xi}\rightarrow\infty$. These fixed points satisfy
\begin{equation}
	c\,\big{|}_{\tilde{\xi}\rightarrow\infty}\, >\,c\,\big{|}_{\tilde{\xi}=1}.
\end{equation}
This relation shows indeed that $\tilde{\xi} \rightarrow \infty$ is a $UV$ fixed point while $\tilde{\xi}=1$ is an $IR$ one. Our motivation for choosing the dual version \eqref{6.3} is intended as a complementary discussion and an additional check of the duality, since in Refs. \cite{Georgiou:2018vbb,Georgiou2020} the $ C $-function was calculated from the action in \eqref{5.18} by using a different method of computation compared to our approach below. Following the Refs. \cite{Shifman} and \cite{Cabra1993}, we place the model in a curved manifold, from which we can obtain correlations of the energy-momentum tensor by taking derivatives of the action with respect to the metric.

As the $C$-function is additive for decoupled theories, we can study the RG flow of the contributions of the partition function \eqref{6.3} separately. The only nontrivial flux is due to the part involving the fields  $g$  and $g^\dagger$. Taking $B=0$ in (\ref{6.3}), we then analyze the RG flow generated by the $SU(N_c)$ piece
\begin{align}
	Z_{SU(N_c)}=\int Dg Dg^\dagger \exp\tilde{k}\left[W[g]+W[g^\dagger]-\frac{\tilde{\xi}}{\pi}\int \dd[2]{z}\Tr \left(g^{\dagger-1}\partial_{\bar{z}}g^\dagger \partial_zgg^{-1}\right)\right]
	\label{sun}
\end{align}
of the total partition function
\begin{eqnarray}
	Z=Z_FZ_gZ_{SU(N_c)},
	\label{3.3}
\end{eqnarray}
with $\tilde{k}=N_f+2N_c$, since $N_f$ is the index of the $N=N_fN_c$ dimensional representation of $SU(N_c)$, and $C_{SU(N_c)}=N_c$.

Going beyond the perturbative regime, this partition function can be evaluated in the large-level limit. To this end, we parametrize the fields as
\begin{equation}
	g^\dagger \equiv  e^{\frac{i}{\sqrt{2\tilde k}}\phi_+}e^{\frac{i}{\sqrt{2\tilde k}}\phi_-} ~~~\text{and}~~~ g \equiv   e^{-\frac{i}{\sqrt{2\tilde k}}\phi_-}   e^{\frac{i}{\sqrt{2\tilde k}}\phi_+}.
	\label{3.1}	
\end{equation}
These parametrizations are convenient as they make manifest the fact that at $\tilde{\xi}=1$ the theory is written in terms of a single field $\phi_+$. According to the above expressions, this implies the decoupling in all orders of the field $\phi_-$, namely, $g^\dagger g = \exp \left( i\sqrt{\frac{2}{\tilde k}}\phi_+\right)$.


\subsection{Leading Order Computation}

The leading order contribution to the effective action is simply 
\begin{align}
	e^{-S_{\text{eff}}^{(0)}}=\int \mathcal{D}\phi_+\mathcal{D}\phi_- \exp -\int \dd[2]{z}\frac{1}{4\pi}\left[\left(1+\tilde{\xi} \right)\phi_+^{a}\nabla \phi_+^{a}+\left(1-\tilde{\xi}\right)\phi_{-}^{a}\nabla\phi_{-}^{a}\right], \label{3.2}
\end{align}
where $ \nabla \equiv \eta^{\mu \nu}\partial_{\mu}\partial_{\nu}$. The corresponding propagators are
\begin{align}
	\expval{\phi^{a}_{\pm}(x)\phi^{b}_{\pm}(y)}=2\pi \frac{\delta^{ab} }{1\pm \tilde{\xi}}G(x,y), \label{3.5}
\end{align}
where $ \nabla G(x,y)=\delta^{(2)}(x-y) $. The singularity at $\tilde{\xi}=1$ is simply a reflex of the sudden decoupling of degrees of freedom that become unphysical due to the emergent gauge invariance \eqref{7.71}. As discussed previously, this leads to a divergence in the partition function that manifests here as a singularity in the propagator of $\phi_-$.  We shall see below how to deal with such behavior.

In addition to the singularity at $\tilde{\xi}=1$, the propagators of massless scalar fields in two dimensions in \eqref{3.5}  famously possess IR divergences that can be controlled through the introduction of mass regulators,  
\begin{align}
	\mathcal{L}_{ct}\sim m_+^{2} \phi_{+}^{a}\phi_+^{a}+m_-^{2}\phi_{-}^{a}\phi_{-}^{a}, \label{3.6}
\end{align}
where $ m_{\pm}^{2} $ are small positive masses, which should be taken to zero at the end of the calculations. 

We shall consider then the properly regularized propagators
\begin{align}
	\expval{\phi_{\pm}^{a}(x)\phi_{\pm}^{b}0)}_{reg}=\mp 2\pi\delta_{a,b}\int \frac{\dd[2]{p}}{(2\pi)^{2}}\frac{e^{i p x}}{(\tilde{\xi}\pm1)p^{2}+m_{\pm}^{2}} =\mp\frac{\delta_{a,b}}{\tilde{\xi}\pm 1 }K_{0}\left(r_{\pm} m_{\pm}\right), 
	\label{3.8.1a}
\end{align}
where $K_n$ is a modified Bessel function of the second kind and we have defined the lengths
\begin{equation}
	r_{\pm} \equiv \frac{|x|}{\sqrt{\tilde{\xi}\pm 1}}.
	\label{3.8.1b}
\end{equation}

Flowing from UV to IR amounts to running $|x|$ from 0 to $L$, with $L$ very large, and accordingly $\tilde{\xi}$ from $\infty$ to $1+\epsilon$, with $\epsilon$ very small. We have introduced the parameters $L$ and $\epsilon$ to keep track of the singularity at $\tilde{\xi}=1$. 
In terms of $r_{\pm}$, this implies 
\begin{equation}
	r_{+}\big{|}_{\text{UV}} =  0 ~~~\text{and}~~~ r_{+}\big{|}_{\text{IR}} =  \frac{L}{\sqrt{2+\epsilon}}\sim\frac{L}{\sqrt{2}},
\end{equation}
and
\begin{equation}
	r_{-}\big{|}_{\text{UV}} =  0 ~~~\text{and}~~~ r_{-}\big{|}_{\text{IR}} =\frac{L}{\sqrt{\epsilon}}.
\end{equation}

As we are working at the leading order, the action is quadratic in the fields and thus the corresponding $C$-function can be obtained using the Noether theorem to compute the energy-momentum tensor following from the action in \eqref{3.2}. The result is
\begin{eqnarray}
	C_{\pm}^{(0)}(r_{\pm})=
	\left(N_{c}^{2}-1\right)\frac{m_{\pm}^4 r_{\pm}^{4}}{4 }\left[\left(K_{2}(m_{\pm} r_{\pm}) \right)^{2}+2 \left(K_{1}(m_{\pm} r_{\pm}\right)^{2}-3\left(K_{0}(m_{\pm} r_{\pm}\right)^{2}\right].
	\label{3.8.1}
\end{eqnarray}

Next we proceed by analyzing the limits $m_{\pm}\rightarrow 0$ and $\epsilon\rightarrow 0$ as we flow from the UV to the IR. To this end, we need to recall the asymptotic behavior of Bessel functions, 
\begin{equation}
	K_n(x) \sim \sqrt{\frac{\pi}{2x}}e^{-x},~~~ x\rightarrow \infty
\end{equation}
and 
\begin{equation}
	K_0(x)\sim -\ln \left(\frac{x}{2}\right)-\gamma~~~\text{and}~~~K_{n>0}(x)\sim \frac{\Gamma(n)}{2}\left(\frac{2}{x}\right)^n, ~~~x\rightarrow 0,
\end{equation}
where $\gamma$ is the Euler–Mascheroni constant. 

With these behaviors, we see that the limit $m_+\rightarrow 0$ can be straightly taken in $C^{(0)}_+$ independent of the limit $\epsilon\rightarrow 0$, which leads to
\begin{equation}
	C^{(0)}_+(r_+)= N_{c}^{2}-1,
	\label{3.8.2a}
\end{equation} 
i.e., it remains constant along the RG flow. This is of course expected for the set of $N_c^2-1$ massless scalar free fields $\phi_+^a$, whose propagator is not singular at $\tilde{\xi}=1$. 

\begin{figure}
	\centering
	\includegraphics[scale=.7,angle=90]{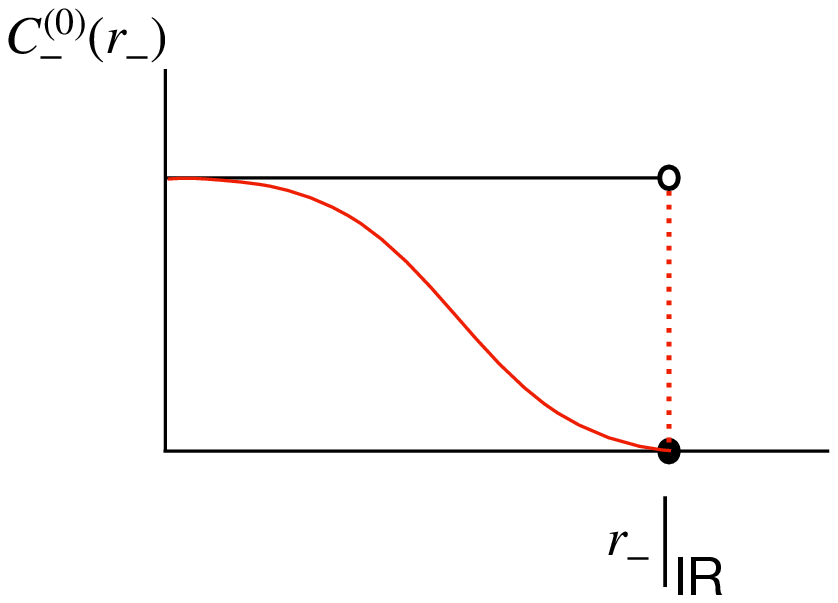}
	\caption{The black line shows the discontinuity of the function $C_{-}^{(0)}(r_{-})$ at $r_{-}\big{|}_{\text{IR}}$. The red line corresponds to the smoothing of the discontinuity due to the mass term.}
	\label{sketch}
\end{figure}

On the other hand, the limit $m_-\rightarrow 0$ in $C^{(0)}_-$ requires some care because $r_{-}\big{|}_{\text{IR}}$ contains an additional divergent factor of $1/\sqrt{\epsilon}$, so that the limits $m_-\rightarrow 0$ and $\epsilon\rightarrow 0$ do not commute at the strict IR. In fact, if we first take $\epsilon\rightarrow 0$, which amounts to $r_{-}\big{|}_{\text{IR}}=\infty$, while keeping $m_-$  finite,
the  $C^{(0)}_-$ function gives
\begin{eqnarray}
	C_{-}^{(0)}\left(r_{-}\big{|}_{\text{IR}}\right)=0.
	\label{3.8.2b}
\end{eqnarray}
Now, along the RG flow, as long as we do not reach the strict IR limit, we can take the limit $m_-\rightarrow 0$, which implies that $C_-^{(0)}(r_-)=N_c^2-1$, except for $r_-=r_{-}\big{|}_{\text{IR}}$. This leads to a discontinuity in the function $C_{-}^{(0)}$, 
\begin{eqnarray}
	C^{(0)}_{-}(r_-)= (N_{c}^{2}-1)\times \left\{\begin{array}{c}
		1 \qquad 0 \leq r_{-} < r_{-}\big{|}_{\text{IR}}, \\ \! 0\qquad ~~~r_{-}=r_{-}\big{|}_{\text{IR}}.
	\end{array}\right. 
\end{eqnarray}
The presence of a small mass $m_-$ along the entire RG flow is a way to smooth the discontinuity of the function $C_{-}^{(0)}$ induced by the singularity at $\tilde{\xi}=1$. Such behaviors are  illustrated in Fig. \ref{sketch}.

Taking into account the contribution of all pieces entering the partition function \eqref{3.3}, the full $C^{(0)}$-function at leading order in the large-$\tilde{k}$ expansion is
\begin{equation}
	C^{(0)}(\tilde{\xi}) = c_{F} + c_{ghost} + C_{+}^{(0)}+ C_{-}^{(0)},  
	\label{cfunction}
\end{equation}
where $c_{F}$ and $c_{ghost}$ are constant terms given by \eqref{7.31} and \eqref{7.32}.

At the fixed points, the $C$-function \eqref{cfunction} leads to the central charges
\begin{align}
	C^{(0)}(\tilde{\xi}\rightarrow \infty)&=\underbrace{N}_{fermions}-\underbrace{2(N_{c}^{2}-1)}_{ghost}+\underbrace{2(N_{c}^{2}-1)}_{WZW}=N\label{3.10.2}\\
	C^{(0)}(\tilde{\xi}=1)&=\underbrace{N}_{fermions}-\underbrace{2(N_{c}^{2}-1)}_{ghost}+\underbrace{(N_{c}^{2}-1)}_{WZW}=N-(N_c^2-1) \label{3.10.1}.
\end{align}
For the ghosts we used $d_{SU(N_c)}=N^2_c-1$ in \eqref{7.32}. These expressions reproduce correctly the central charge of free fermions at the UV, and the central charge of \eqref{7.9} at the IR in the large-$\tilde{k}$ expansion.


\subsection{Next Order Computation}

Now we move to the computation of the $C$-function up to the subleading order $1/\tilde k$. As the corresponding contributions to the effective action are not quadratic in the fields, it is no longer convenient to proceed like we did in the previous case. 

The strategy consists of placing the theory in a curved background with metric $\gamma_{\mu\nu}$, so that the classical energy-momentum tensor can be computed as
\begin{equation}
	T_{\mu\nu}(x)=-\frac{2}{\sqrt{\gamma}}\frac{\delta S}{\delta \gamma^{\mu\nu}(x)}.
\end{equation}
Then, from the effective action
\begin{equation}
	e^{-S_{\text{eff}}[\gamma]}\equiv\int \mathcal{D}g \mathcal{D}g^\dagger e^{-S[\gamma,g,g^\dagger]},
\end{equation}
we obtain the correlation functions of the energy-momentum tensor by taking functional derivatives with respect to the metric. In particular, the two-point function reads
\begin{eqnarray}
	-\frac{2}{\sqrt{\gamma}(x)}\frac{2}{\sqrt{\gamma}(y)}\frac{\delta^2 S_{\text{eff}}[\gamma]}{\delta \gamma^{\mu\nu}(x) \delta \gamma^{\rho\sigma}(y)}&=& \langle T_{\mu\nu}(x)T_{\rho\sigma}(y)\rangle-\langle T_{\mu\nu}(x)\rangle\langle T_{\rho\sigma}(y)\rangle\nonumber\\
	&-&\frac{2}{\sqrt{\gamma}(x)}\frac{2}{\sqrt{\gamma}(y)}\left\langle\frac{\delta^2 S}{\delta \gamma^{\mu\nu}(x) \delta \gamma^{\rho\sigma}(y)}\right\rangle.\label{tt}
\end{eqnarray}
The $C$-function can be obtained from this relation as we take the flat space limit. The one-point functions of the energy-momentum tensor do not contribute in the flat space limit, whereas the 
term in the second line leads just to a contact term and also does not contribute to the $C$-function, which is defined at separated points. Therefore, the $C$-function reduces essentially to a combination of second derivatives of the effective action.

In the large-$\tilde k$ expansion, the correction of order $1/\tilde k$ comes from
\begin{align}
	S^{(1/\tilde k)}_{\text{eff}}[\gamma]=\expval{S_{4}}-\frac{1}{2}\expval{S_{3}S_{3}}\label{3.11},
\end{align}
where $ S_{3} $ and $S_{4}$ are the three and four fields terms of the action in \eqref{sun} with the expansion \eqref{3.1}. 

Before writing the theory in a curved background, it is convenient to proceed momentarily in the flat space to obtain explicit forms of $S_3$ and $S_4$ in terms of $\phi_{+}$ and $\phi_-$,
\begin{align}
	S_{3}&=-\frac{1}{12\pi \sqrt{2\tilde k}}\int \dd[2]{z} f_{abc} \left[i \epsilon_{\mu \nu} \left(3(1-\tilde{\xi})\phi_+^{a}\partial_{\mu}\phi_-^{b}\partial_{\nu}\phi_-^{c}+(1+3\tilde{\xi})\phi_+^{a}\partial_{\mu}\phi_+^{b}\partial_{\nu}\phi_+^{c}\right)\right.\nonumber \\
	&+\left.3\eta^{\mu \nu}(1-\tilde{\xi})\phi_+^{a}\partial_{\mu}\phi_+^{b}\partial_{\nu}\phi_-^{c}\right] \label{S3}
\end{align}	
and	
\begin{align}
	S_{4}&=-\frac{1}{96\pi \tilde k}\int\dd[2]{z} f_{abe}f_{cde}\eta^{\mu \nu}\left[(1-\tilde{\xi})\phi_-^{a}\partial_{\mu}\phi_-^{b}\partial_{\nu}\phi_-^{c}\phi_-^{d}+6(1-\tilde{\xi})\phi_-^{a}\partial_{\mu}\phi_-^{b}\partial_{\nu}\phi_{+}^{c}\phi_+^{d}\right.\nonumber\\
	&+\left.  (1 + 7\tilde{\xi})\phi_+^{a}\partial_{\mu}\phi_+^{b}\partial_{\nu}\phi_+^{c}\phi_+^{d}\right]\label{S4}.
\end{align}
In $S_4$ we have neglected terms of the type $ \epsilon^{\mu \nu} \phi_+ \partial_{\mu}\phi_+\partial_{\nu}\phi_- \phi_+$, which do not contribute as there is no correlation $ \expval{\phi_+\phi_-}$. This type of term only contributes in higher orders in the large-$\tilde k$ expansion.

Now we place the theory in the curved background, which means to introduce the metric dependence in $S_3[\gamma]$ and $S_4[\gamma]$. Then, taking the expected values we obtain the contribution of order $1/\tilde k$ to the effective action
\begin{align}
	S_{\text{eff}}^{(1/\tilde k)}[\gamma]&=\frac{\pi N_{c}(N_{c}^{2}-1)}{3 \tilde k}\frac{\left(3\tilde{\xi}^2-4\tilde{\xi}-1\right)}{(\tilde{\xi} +1)^2 (1-\tilde{\xi})}\int \dd[2]{x} \sqrt{\gamma}\gamma^{\mu \nu}\left[\partial_{\mu}^{x}G \partial_{\nu}^{y}G -G  \partial_{\mu}^{x}\partial_{\nu}^{y}G \right]_{x=y}\label{3.12}\\
	&-\frac{4\pi N_{c}(N_{c}^{2}-1)}{3\tilde k}\frac{ \left(3 \tilde{\xi}^2+3\tilde{\xi} +1\right)}{(1+\tilde{\xi})^3}\int \dd[2]{x}\dd[2]{y}\epsilon^{\mu \nu}\epsilon^{\sigma\rho} G  \partial_{\mu}^{x}\partial_{\sigma}^{y}G \partial_{\nu}^{x}\partial_{\rho}^{y}G  \nonumber \\
	&+\frac{\pi \left(1-\tilde{\xi}\right)N_{c}(N_{c}^{2}-1)}{2\tilde k(1+\tilde{\xi})^{2}}\int \dd[2]{x}\dd[2]{y}\gamma^{\frac{1}{2}}(x)\gamma^{\frac{1}{2}}(y)\gamma^{\mu \nu}(x)\gamma^{\sigma\rho}(y)\left[G \partial_{\mu}^{x}\partial_\sigma^{y}G  \partial_{\nu}^{x}\partial_{\rho}^{y}G\right.\nonumber\\ &- \left. \partial_{\mu}^{x}G  \partial_{\sigma}^{y}G \partial_{\nu}^{x}\partial_{\rho}^{y}G \right]\nonumber, 
\end{align}
where $ G $ is the propagator in the curved spacetime, namely, 
\begin{equation}
	\frac{1}{\sqrt{\gamma}} \partial_{\mu}(\sqrt{\gamma}\gamma^{\mu\nu}\partial_{\nu}G(x,y))=\frac{1}{\sqrt{\gamma}} \delta^{(2)}(x-y).
\end{equation}

All the integrals appearing in \eqref{3.12}  can be expressed in terms of the Polyakov action 
\begin{align}
	\Gamma[\gamma]\equiv\frac{1}{96\pi} \int \dd[2]{x} \dd[2]{y} \sqrt{\gamma(x)}\sqrt{\gamma(y)} R(x)G(x,y) R(y) .
\end{align}
The manipulations involved in these computations are intricate so that we relegate the details to the  Appendix \ref{AC}. The resulting effective action reads 
\begin{align}
	S_{\text{eff}}^{(1/\tilde k)}[\gamma]=\frac{N_{c}(N_{c}^{2}-1)}{\tilde k} f(\tilde{\xi}) \Gamma[\gamma],  \label{3.15}
\end{align}
where 
\begin{align}
	f(\tilde{\xi})&\equiv  \frac{2(2\tilde{\xi}+1)}{(1-\tilde{\xi})(1+\tilde{\xi})^{3}}.
	\label{3.16}
\end{align}

Employing \eqref{tt}, we can compute the two-point function of the energy-momentum tensor simply by taking derivatives of $S_{\text{eff}}[\gamma]$ with respect to the metric and then taking the flat space limit. This leads to 
\begin{align}
	-\frac{2}{\sqrt{\gamma}(x)}\frac{2}{\sqrt{\gamma}(y)}\frac{\delta^2 S_{\text{eff}}[\gamma]}{\delta \gamma^{\mu\nu}(x) \delta \gamma^{\rho\sigma}(y)}\Big{|}_{\gamma=\eta}&=-\frac{1}{12\pi}\left[\frac{1}{2\pi}\partial_{\mu}\partial_{\nu}\partial_{\sigma}\partial_{\rho}\ln \abs{x-y}\right. \label{3.9.1} \\ &- \left(\eta_{\mu \nu} \partial_{\sigma}\partial_{\rho}+\eta_{\sigma\rho}\partial_{\mu }\partial_{\nu}\right)\delta^{2}(x-y)+\eta_{\mu \nu}\eta_{\sigma\rho}\partial^{2}\delta^{2}(x-y)\Big]\nonumber. 
\end{align} 
Picking specific indices, we can find the components entering the $C$-function \eqref{3.0}, i.e.,    
\begin{align}
	\expval{T(x)T(0)}^{(1/\tilde k)}&= \frac{N_{c}(N_{c}^{2}-1)}{8 \pi^{2}\tilde k} f(\tilde{\xi})\frac{1}{x^{4}}, \nonumber\\
	\expval{T(x) \Theta(0)}^{(1/\tilde k)}&=0, \nonumber\\
	\expval{\Theta(x) \Theta(0)}^{(1/\tilde k)}&=0. \label{3.10}
\end{align}
From these relations we can immediately read the contribution of order $1/\tilde k$ to the $C$-function,
\begin{align}
	C^{(1/\tilde k)}(\tilde{\xi})=\left(N_{c}^{2}-1\right)\frac{N_{c}}{\tilde k} \frac{2(2\tilde{\xi}+1)}{(1-\tilde{\xi})(1+\tilde{\xi})^{3}}.
	\label{3.17}
\end{align}
There are several interesting points to notice in this expression. First, we see that both $C^{(1/\tilde k)}$ and $\partial_{\tilde{\xi}}C^{(1/\tilde k)}$ vanish at the UV limit $\tilde{\xi}\rightarrow\infty$. This is a suitable behavior of the $C$-function since it must be stationary at the fixed points and the zero order contribution \eqref{3.10.2} already exhausts the free fermion central charge. Furthermore, $C^{(1/\tilde k)}$  approaches to zero by negative values, ensuring that  $C^{(0)}+C^{(1/\tilde k)}< C_{UV}$ is always satisfied along the RG flow. On the other hand, the divergent behavior in the limit $\tilde{\xi}\rightarrow 1$ is even more dramatic than the discontinuous zero order contribution. So the main question is how close to the IR fixed point can we reach through RG flow? Relatedly, since \eqref{3.17} is negative for $\tilde{\xi}$ slightly larger than one, this seems to violate the inequality $C_{IR}<C^{(0)}+C^{(1/\tilde k)}$.

To address these issues, we notice that our large-$\tilde k$ computation of the $C$-function can be organized in an expansion of the form
\begin{equation}
	C(\tilde{\xi})=\sum_{n=0}^{\infty}\left(\frac{N_c}{\tilde k}\frac{1}{1-\tilde{\xi}}\right)^n f_n(\tilde{\xi}),
	\label{cf}
\end{equation}
where $f_n(\tilde{\xi})$ is a smooth function at $\tilde{\xi}=1$ and $f_n(\tilde{\xi}\rightarrow\infty)\rightarrow 0$ for $n>0$. Comparing with the previous computations we see that
\begin{equation}
	f_0=N~~~\text{and}~~~ f_1=(N_c^2-1)\frac{2(2\tilde{\xi}+1)}{(1+\tilde{\xi})^{3}}.
\end{equation}

Now, the crucial point of \eqref{cf} is that we can probe the IR limit if we consider at the same time that $\tilde k$ is large enough, since the divergence in $\tilde{\xi}\rightarrow1$ can be balanced for large values of $\tilde k$. This can be implemented in a systematic way through the zoom-in limit \cite{Sfetsos:2013wia,Georgiou2015,Georgiou2017,georgiou2017integrable,Georgiou2017a,Georgiou:2018vbb,hoare2019integrable,hoare2019integrable,Georgiou2020}, in which we link the two limits $\tilde{\xi}\rightarrow1$ and $\tilde k\rightarrow\infty$ in the IR region,
\begin{equation}
	\tilde{\xi}=1 + \frac{\epsilon}{\tilde k},
	\label{z}
\end{equation} 
with $\epsilon$  being a positive parameter and $\epsilon\ll \tilde k$. With this, the $C$-function in \eqref{cf} can be expressed in terms of a double expansion
\begin{equation}
	C(\epsilon)=\sum_{n=0}^{\infty}\sum_{m=0}^{\infty} \alpha_{nm}(-1)^n\left(\frac{N_c}{\epsilon}\right)^n \left(\frac{\epsilon}{\tilde k}\right)^m,
	\label{de}
\end{equation}
where presumably 
\begin{equation}
	\frac{N_c}{\epsilon}\lesssim 1.
\end{equation}
Therefore, we need
\begin{equation}
	N_c \lesssim \epsilon \ll 2N_c + N_f,
\end{equation}
which can be easily satisfied simply by considering $N_f \gg N_c$.

With these considerations it is interesting to compare the IR behavior of the $C$-function in terms of the double expansion \eqref{de}, 
\begin{equation}
	C(\epsilon) = N -(N_c^2-1)\left(\frac{N_c}{\epsilon}\right) \left(\frac{3}{4}-\frac{5}{8}\frac{\epsilon}{\tilde k}+ \frac{3}{8} \left(\frac{\epsilon}{\tilde k}\right)^2+\cdots\right)+\cdots,
\end{equation}
with the large-$\tilde{k}$ limit of the IR central charge, given by 
\begin{equation}
	c_{IR} = N -(N_c^2-1)+\cdots.
\end{equation}
Thus, no matter how close to $\tilde{\xi}=1$ we are able to reach through the limit of $\tilde k\rightarrow \infty$, there is a discontinuity in the $C$-function, which is a reflex of the emergent gauge invariance at $\tilde{\xi}=1$. Physically, this means that some physical degrees of freedom are suddenly decoupled from the model.


\subsection{$\beta$-function}

The $C$-function can also be used to compute the RG $\beta$-function according to  
\begin{align}
	\partial_{\tilde{\xi}}C(\tilde{\xi})=24 g_{\tilde{\xi},\tilde{\xi}}\beta_{\tilde{\xi}},
	\label{ur}
\end{align}
where $g_{\tilde{\xi},\tilde{\xi}}$ is the Zamolodchikov metric associated with current-current interactions \cite{Zomolodchikov1986}. For such case, the Abelian part of the metric is  independent of $\tilde k$ and was first calculated in \cite{Kutasov1989}, but a more recent discussion can also be found in the Appendix of \cite{Georgiou2015}. The result is
\begin{align}
	g_{\tilde{\xi},\tilde{\xi}}=\frac12\frac{(N_c^2-1)}{(1-\tilde{\xi}^{2})^{2}}.
	\label{zm}
\end{align}
An immediate problem in using this metric to compute the $\beta$-function is that it vanishes at the UV fixed point $\tilde{\xi}=\infty$. To circumvent this issue we can use the duality transformations \eqref{6.4} and \eqref{6.5}. With these, the UV fixed point $\tilde{\xi}=\infty$ is mapped to the origin ${\xi}=0$, whereas the IR fixed-point $\tilde{\xi}=1$ is self-dual.
Using the duality in \eqref{ur}, we obtain
\begin{align}
	\beta(\xi)=-\frac{N_{c}}{ k}\frac{\xi^{2}}{(1+\xi)^{2}}.
	\label{beta}
\end{align}
This result agrees with the $\beta$-function computed in \cite{Georgiou2020}. Note that the $\beta$-function \eqref{beta} has the trivial fixed point ${\xi}=0$, whereas the value ${\xi}=1$ does not show up as a fixed point. 

\begin{figure}
	\centering
	\includegraphics[scale=.7,angle=90]{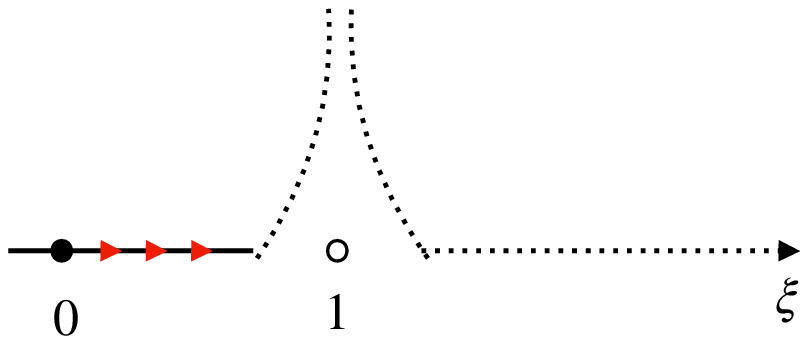}
	\caption{The parameter space with the fixed points. We represent the unattainability of the IR fixed point $\xi=1$ as a point outside the one-dimensional parameter space.}
	\label{phase}
\end{figure}

It is interesting to analyze the IR fixed point from the perspective of a geodesic distance in the parameter space, which is defined as \cite{Kutasov1989}
\begin{align}
	\Delta S\equiv \int_{{\xi}_1}^{{\xi}_2} \sqrt{g_{{\xi},{\xi}}}\dd{{\xi}}=\left.\sqrt{2 (N_{c}^{2}-1)}\log\frac{1+{\xi}}{1-{\xi}}\,\right|_{{\xi}_1}^{{\xi}_2}.
\end{align} 
We see that the IR fixed point ${\xi}=1$ is infinitely far apart from any point of the parameter space. This is illustrated in Fig. \ref{phase}.

\subsection{Deformation and Emergent Gauge Invariance}

With a slightly deformation of the theory we can follow closely the fate of the IR fixed point and understand better its peculiar behavior. Consider the action
\begin{align}
	S[g,g^\dagger;\tilde k_1,\tilde k_2]&=-\tilde k_{1}W[g^\dagger]-\tilde k_{2}W[g]+\frac{\tilde{\xi} \sqrt{\tilde k_{1}\tilde k_{2}}}{\pi}\!\!\!\int \!\! \dd[2]{x}\text{Tr} g^{\dagger -1}\partial_{\bar{z}}g^\dagger \partial_zgg^{-1}\nonumber\\
	&=-\tilde k_2 \left[\tilde{\xi}_0^2 W[g^\dagger] + W[g] - \frac{\tilde{\xi} \tilde{\xi}_{0}}{\pi}\!\!\!\int \!\! \dd[2]{x}\text{Tr} g^{\dagger-1}\partial_{\bar{z}}g^\dagger \partial_zgg^{-1}\right],
	\label{def}
\end{align}
where we have defined $\tilde{\xi}_{0} \equiv \sqrt{\tilde k_1/\tilde k_2}$. Setting $\tilde k_1=\tilde k_2 \,(\tilde{\xi}_{0}=1)$ we recover our original theory \eqref{sun}. The action \eqref{def} possesses a generalized version of the duality \eqref{6.4} and \eqref{6.5}. In the large-$\tilde k$ limit, it reads \cite{georgiou2017integrable}
\begin{equation}
	\tilde k_{1}\rightarrow -\tilde k_{2},~~~ \tilde k_{2}\rightarrow -\tilde k_{1},~~~ \text{and} ~~~ \tilde{\xi}\rightarrow \frac{1}{\tilde{\xi}}.
	\label{T-duality}
\end{equation}

The theory \eqref{def} has two fixed points at finite $\tilde{\xi}$, namely, $\tilde{\xi}=\tilde{\xi}_{0}$ and $\tilde{\xi}=\tilde{\xi}_{0}^{-1}$. The action in these cases acquires the respective forms
\begin{equation}
	\left.S[g,g^\dagger;\tilde k_1,\tilde k_2]\right|_{\tilde{\xi}=\tilde{\xi}_{0}}= -\tilde k_1 W[g^\dagger g] - (\tilde k_2-\tilde k_1) W[g]
\end{equation}
and 
\begin{equation}
	\left.S[g,g^\dagger;\tilde k_1,\tilde k_2]\right|_{\tilde{\xi}=\tilde{\xi}_{0}^{-1}}= -\tilde k_2 W[g^\dagger g] - (\tilde k_1-\tilde k_2) W[g^\dagger],
\end{equation}
where we have used the Polyakov-Wiegmann identity \eqref{5.10}. It is interesting to compare these actions with $W[g^\dagger g]$. Notice that, in contrast to $W[g^\dagger g]$, the above actions do not have the emergent gauge invariance \eqref{7.71}, except when $\tilde k_1=\tilde k_2$. Therefore, the theory with $\tilde k_1\neq \tilde k_2$ should not be afflicted with the singular behavior at the fixed points $\tilde{\xi}_0$ and $\tilde{\xi}_{0}^{-1}$. 

We can see this quite explicitly through the $\beta$-function, which has been computed in \cite{Georgiou2017} for the dual theory with different levels,   
\begin{equation}
	\beta({\xi})=-\frac{N_{c}}{\sqrt{k_{1}k_{2}}}\frac{{\xi}^{2}({\xi}-{\xi}_{0})({\xi}-{\xi}_{0}^{-1})}{(1-{\xi}^{2})^{2}}.
\end{equation}
It shows explicitly the existence of fixed points at ${\xi}_{0}$ and ${\xi}_{0}^{-1}$. Now we can see that they disappear when ${\xi}_{0}\rightarrow 1$  and ${\xi}_{0}^{-1}\rightarrow1$. In fact, in this situation the $\beta$-function reduces to  
\begin{align}
	\lim \limits_{{\xi}_{0}\rightarrow 1}\beta({\xi})=-\frac{N_{c}}{k}\frac{{\xi}^{2}}{(1+{\xi})^{2}}
\end{align}
which is precisely the result in \eqref{beta}.


\section{Final Remarks} \label{Section6}

Throughout this work we have examined several aspects of the non-Abelian Thirring model, with the main purpose of studying its RG properties. One of the most direct ways to find the fixed points is through bosonization, as they appear explicitly in the bosonic theory.  The bosonization route however is not unique, so that using suitable combinations of the Polyakov-Wiegman identity, we are able to derive distintic, but equivalent, bosonic forms of the model. These are interesting in their own since they enjoy a remarkable strong/weak duality, and are the starting point of the computation of the $C$-function in the large-level limit.

The RG flow of the non-Abelian Thirring model possesses a quite peculiar feature: the IR fixed point is strictly unreachable through RG flow due to an emergent gauge invariance. This property has interesting and unusual consequences. Specifically, it leads to a discontinuity in the $C$-function, making the study of the flow to the IR subtle. We discuss this point carefully from the perspective of the dual theory, and show that we can reach close enough to the IR fixed point in the large-level limit. Indeed, this produces a two-parameter expansion, which becomes more accurate as $N_f \gg N_c$.

Our main motivation for this work is to employ it in the construction of topological phases of matter in the framework of quantum wires. In that context, we usually start with a set of one-dimensional free fermions, and then introduce interactions both along as well as between neighboring wires. This sort of microscopic model is able to stabilize a two-dimensional spatial phase if the interactions are relevant in the bulk, opening a gap in the spectrum. Therefore, understanding the fate of the fermionic theory in the IR is a crucial ingredient in this construction. Fermions with an internal group structure interacting through current-current operators are the basic building-blocks of this approach, and can be modeled in terms of  the non-Abelian Thirring theory. We address this problem in a separate publication \cite{Corso:2023}.

\appendix
\section{Effective Action Regularization}\label{AC}

Here we want to discuss the procedure to regularize the effective action \eqref{3.12}. This section is intended to familiarize the reader with the methods we used in a convenient way, and is largely based in previous works, namely \cite{Shifman,francesco}. Before we do that, it is convenient to review the free boson effective action
\begin{align}
	\Gamma=\frac{1}{2}\tr \ln-\nabla,\label{b1}
\end{align}
where $ \nabla=\frac{1}{\sqrt{\gamma}}\partial_{\nu}\left(\sqrt{\gamma}\gamma^{\mu \nu}\partial_{\nu}\right) $ is the Laplace-Beltrami operator. We start by giving an integral representation to the logarithm,
\begin{align}
	\ln x=-\int\limits_{\epsilon}^{\infty}\frac{\dd{t}}{t}\left[e^{-x t}-e^{-t}\right], \label{b2}
\end{align}
where the limit $ \epsilon\rightarrow 0 $ must be taken in the end of the calculations. Under an infinitesimal scale transformation of the metric, $ \gamma_{\mu \nu}^{\prime}=e^{\delta \omega} \;\gamma_{\mu \nu} $, the Laplace-Beltrami operator transforms as $\nabla^{\prime}=e^{-\delta \omega}\nabla$ and we can write the variation of the boson effective action as
\begin{align}
	\delta \Gamma&=\frac{1}{2}\tr \int \limits_{\epsilon}^{\infty}\dd{t}\delta \omega \nabla e^{t \nabla}=\frac{1}{2}\tr \int \limits_{\epsilon}^{\infty}\dd{t}\delta \omega \frac{d}{dt}e^{t \nabla}=-\frac{1}{2}\tr \delta\omega e^{\epsilon \nabla}\nonumber\\
	\delta \Gamma&=-\frac{1}{2}\int \dd^{2}x \;\sqrt{\gamma}\delta \omega(x) K(x,x,\epsilon), \label{b3}
\end{align}
where $ K(x,x,\epsilon) $ is the heat kernel of the Laplace-Beltrami operator
\begin{align}
	K(x,y,t)=\mel{x}{e^{t \nabla}}{y}, \quad \text{for} \quad t \geq 0. \label{b4}
\end{align}

In the limit $ \epsilon \rightarrow 0 $, we can expand the heat kernel in powers of the infinitesimal time
\begin{align}
	\delta \Gamma=-\frac{1}{2}\int \dd[2]{x} \sqrt{\gamma}\delta \omega(x) \left(\frac{1}{4\pi \epsilon}+\frac{R(x)}{24\pi}\right), \label{b5}
\end{align}
where $ R(x) $ is the scalar curvature. The divergent term inside the bracket can be traced back to our assumption that the manifold is finite, it has nothing to do with curvature. In this way, the divergence can be eliminated by the addition of a local, field independent, counterterm to the effective action. Now we integrate the effective action, as the metric changes $ \gamma_{\mu \nu}^{\prime}=e^{\omega} \gamma_{\mu \nu} $ so does the Ricci scalar $ R^{\prime}=e^{-\omega}\left(R-\nabla \omega\right) $, such that the effective action 
\begin{align}
	\Gamma=-\frac{1}{96\pi}\int \dd^{2}x\; \sqrt{\gamma} \left(\partial_{\mu} \omega\partial^{\mu}\omega+2 R \omega \right),\label{b6}
\end{align}
which can be checked to be equivalent to the usual free boson effective action \cite{POLYAKOV1981207}, 
\begin{align}
	\Gamma=\frac{1}{96\pi}\int \dd^{2}x \dd^{2}y \; \sqrt{\gamma(x)}\sqrt{\gamma(y)}R(x)G(x,y)R(y). \label{b7}
\end{align}


At this point we can relate the integrals in the two loops effective action \eqref{3.12} to the free boson effective action \eqref{b6}, from which, we can derive the $ C $-function by taking metric derivatives.  Let us start with the first term
\begin{align}
	I_{1}=\int \dd^{2}x \; \sqrt{\gamma}\gamma^{\mu \nu}\left[\partial_{\mu}^{x}G\partial_{\nu}^{y}G-G\partial_{\mu}^{x}\partial_{\nu}^{y}G\right]_{x=y}.\label{b8}
\end{align}

We start by separating the finite part of the propagator $ \bar G $
\begin{align}
	\bar{G}(x,y)&\equiv-\frac{1}{2\pi}\ln s(x,y)+G(x,y),
\end{align}
such that at coinciding points
\begin{align}
	G(x,x)=\bar G(x,x)+c\equiv\bar{G}(x)+c,
\end{align}
where $ s(x,y) $ is the geodesic distance between the points $ x $ and $ y $ and $ c\propto (d-2)^{-1} $  is a constant. The logarithm term completely accounts for the divergence that is present in the propagator, such that $ \bar{G}(x,y) $ has no divergence. In the limit $ x=y $ we regularize the divergence, in doing so $ G(x,y)\rightarrow\bar{G}(x,y) $. Furthermore, we can use the symmetry of the propagator to show that 
\begin{align}
	\eval{	\partial_{\mu}^{x}G(x,y)}_{x=y}=\eval{\partial_{\mu}^{y}G(x,y)}_{x=y}=\frac{1}{2}\partial_{\mu}^{x}\bar{G}(x),\label{b10}
\end{align}  
At last, we need the identity 
\begin{align}
	\nabla \bar{G}(x)=2\left[\nabla G(x,y)+\partial_{\mu}^{x}\partial^{ \mu,y}G(x,y)\right]_{x=y} \label{b11},
\end{align}
from which we derive 
\begin{align}
	\eval{\partial_{x_\mu}\partial^{y_\mu}G(x,y)}_{x=y}=\frac{R}{8\pi}.
\end{align}

Using the identities we rewrite
\begin{align}
	I_{1}&=\int \dd[2]{x} \sqrt{\gamma}\left[\frac{1}{4}\partial_{\mu}\bar G\partial^{\mu}\bar G-\frac{1}{8\pi}\bar GR\right]=-\frac{1}{4}\int\dd[2]{x}\sqrt{\gamma} \left[\bar{G}\nabla \bar{G}+\frac{1}{2\pi}\bar{G}R\right]\nonumber\\
	&=-\frac{3}{16\pi}\int \dd[2]{x} \bar G R.
\end{align}
Let us recall that we are looking for correlations functions of the energy momentum tensor, which can be obtained by extracting the metric dependence of the effective action. To this end, we consider a conformal tranformation of the metric $ \gamma_{\mu \nu}\rightarrow \gamma_{\mu \nu}^{\prime}=e^{\omega}\gamma_{\mu \nu} $, and look for how $ I_{1} $ changes under such transformation. As expected, the Green function and the Ricci scalar are sensitive to this transformation,
\begin{align}
	\bar{G}^{\prime}(x)=\bar{G}(x)-\frac{\omega(x)}{4\pi}+\cdots \qquad \text{and }\qquad R^{\prime}=e^{-\omega}(R-\nabla \omega), \label{b12}
\end{align}
such that the integral undergoes the change
\begin{align}
	\delta_{\omega}I_{1}=\frac{3}{16\pi}\int \dd[2]{x}\sqrt{\gamma} \left[\frac{1}{4\pi}\partial_{\mu}\omega\partial^{\mu}\omega+\frac{\omega R}{4\pi}+\omega\nabla \bar G\right]=\frac{3}{64\pi^{2}}\int \dd[2]{x} \sqrt{\gamma}\left[\partial_{\mu}\omega\partial^{\mu}\omega+2R \omega\right].
\end{align}
Comparing this with the free boson effective action, equation \eqref{b6}, leads us to 
\begin{align}
	I_{1} \approx -\frac{9}{2\pi}\Gamma,
\end{align}
where the symbol $ \approx $ is meat to signify that the two sides have the same metric dependence.

The last integral in the effective action \eqref{3.12}
\begin{align}
	I_{3}=\int \dd[2]{x}\dd[2]{y}\sqrt{\gamma(x)}\sqrt{\gamma(y)}\gamma^{\mu \nu}(x)\gamma^{\sigma\rho}(y)\Big[\underbrace{G \partial_{\mu}^{x}\partial_\sigma^{y}G  \partial_{\nu}^{x}\partial_{\rho}^{y}G}_{A} -\underbrace{\partial_{\mu}^{x}G  \partial_{\sigma}^{y}G \partial_{\nu}^{x}\partial_{\rho}^{y}G}_{B} \Big]=A-B \label{b30}
\end{align}
can be related to the individual terms of the integral $ I_{1} $ through partial integrations and the propagator equation. We start by considering 
\begin{align}
	B&\equiv\int \dd[2]{x}\dd[2]{y}\sqrt{\gamma(x)}\sqrt{\gamma(y)}\gamma^{\mu \nu}(x)\gamma^{\sigma\rho}(y)\partial_{\mu}^{x}G  \partial_{\sigma}^{y}G \partial_{\nu}^{x}\partial_{\rho}^{y}G\nonumber\\
	&=-\int \dd[2]{x}\dd[2]{y}\sqrt{\gamma(x)}\sqrt{\gamma(y)}\gamma^{\mu \nu}(x)\gamma^{\sigma\rho}(y)\left[\partial_{\mu}^{x}G\nabla^{y} G\partial_{\nu}^{x}G+\partial_{\rho}^{y}\partial_{\mu}^{x}G\partial_{\sigma}^{y}G\partial_{\nu}^{x}G\right]. \label{b31}
\end{align}
Notice that the last term in the integral equals $ B $ under the index change $ \mu \leftrightarrow \nu $, thus we write
\begin{align}
	B&=-\frac{1}{2}\int \dd[2]{x}\dd[2]{y}\sqrt{\gamma(x)}\sqrt{\gamma(x)}\gamma^{\mu \nu}\partial_{\mu}^{x}G\partial_{\nu}^{x}G \nabla^{y}G\nonumber\\
	&=-\frac{1}{2}\int \dd[2]{x} \dd[2]{y} \sqrt{\gamma(x)}\sqrt{\gamma(y)}\gamma^{\mu \nu}\partial_{\mu}^{x}G\partial_{\nu}^{x}G \delta^{(2)}(x-y) \nonumber\\
	&=-\frac{1}{2}\int \dd[2]{x} \sqrt{\gamma}\gamma^{\mu \nu}\eval{\partial_{\mu}G\partial_{\nu}G}_{x=y}=-\frac{3}{4\pi}\Gamma. \label{b32}
\end{align} 
Applying similar methods in $ A $ yields
\begin{align}
	A&\equiv\int \dd[2]{x}\dd[2]{y} \sqrt{\gamma(x)}\sqrt{\gamma(y)}\gamma^{\mu \nu}\gamma^{\sigma\rho}G \partial_{\mu}^{x}\partial_\sigma^{y}G  \partial_{\nu}^{x}\partial_{\rho}^{y}G \nonumber\\
	&=-\int \dd[2]{x} \dd[2]{y}\sqrt{\gamma}\sqrt{\gamma}\gamma^{\mu \nu}
	\left[\partial_{\rho}^{y}G\partial_{\mu}^{x}\partial_{\sigma}^{y}G\partial_{\nu}^{x}G+G \partial_{\mu}^{x}\nabla^{y}G\partial_{\nu}^{x}G\right]\nonumber\\
	&=-B+\int \dd[2]{x}\sqrt{\gamma(x)}\gamma^{\mu \nu}(x)\left[\partial_{\mu}^{x}G\partial_{\nu}^{x}G+G \nabla^{x}G\right]_{x=y}=\frac{9}{4\pi}\Gamma \label{b33}
\end{align}
which also follows from the discussions of $ I_{1} $. At last, we return to the action \eqref{b30} to find
\begin{align}
	I_{3}=A-B=\frac{3}{\pi}\Gamma.
\end{align}

The remaining term involves a much more complicated and lengthy calculation, which is beyond the scope of this work. Here we reproduce the result of \cite{Shifman}, which reads
\begin{align}
	I_{2}&=\int \dd[2]{x}\dd[2]{y}\epsilon^{\mu \nu}\epsilon^{\sigma\rho} G(x,y) \partial_{\mu}^{x}\partial_{\sigma}^{y}G(x,y)\partial_{\nu}^{x}\partial_{\rho}^{y}G(x,y)=\frac{3}{4\pi}\Gamma.
\end{align}

\acknowledgments

This work is partially supported by Brazilian agencies CAPES and  CNPq.



\bibliographystyle{ieeetr}
\bibliography{thesis}

%
%
%
%
%
%
%
%
\end{document}